\documentclass[aps,prb,floatfix,twocolumn,superscriptaddress]{revtex4-2}
\usepackage{graphicx}
\usepackage[english]{babel}
\usepackage{amsmath}
\usepackage{amssymb}
\usepackage{tensor}

\newcommand{\vk}{\mathbf{k}}

\newcommand{\be}{\begin{eqnarray}}
\newcommand{\ee}{\end{eqnarray}}
\newcommand{\p}{\partial}

\newcommand{\dc}{c^{\dagger}}

\def\ket#1{|#1\rangle}
\def\bra#1{\langle #1 |}
\def\ep#1{\langle #1 \rangle}

\begin{document}

\title{Uhlmann holonomy against Lindblad dynamics of topological systems at finite temperatures}

\author{Yan He}
\affiliation{College of Physics, Sichuan University, Chengdu, Sichuan 610064, China}
\email{heyan$_$ctp@scu.edu.cn}

\author{Chih-Chun Chien}
\affiliation{Department of Physics, University of California, Merced, CA 95343, USA.}
\email{cchien5@ucmerced.edu}

\begin{abstract}
The Uhlmann phase, which reflects the holonomy as the purified state of a density matrix traverses a loop in the parameter space, has been used to characterize topological properties of several systems at finite temperatures. We test the Uhlmann holonomy against various system-environment couplings in quantum dynamics described by the Lindblad equations of three topological systems, including the Su-Schrieffer-Heeger (SSH) model, Kitaev chain, and Bernevig-Hughes-Zhang (BHZ) model. The Uhlmann phase is shown to remain quantized in all the examples if the initial state is topological and only certain types of the Lindblad jump operators are present. Topological protection at finite temperatures against environmental effects in quantum dynamics is therefore demonstrated albeit only for a restricted class of system-environment couplings.
\end{abstract}

\maketitle

\section{Introduction}
Holonomy is a geometric concept that measures the change of a horizontal-lifted curve over a loop in the base space~\cite{Bohm_book,Nakahara}. The Berry phase provides a physical realization of the Berry holonomy of pure quantum states by reflecting the phase change of the wavefunction as the system traverses a loop in the parameter space~\cite{Berry}. For one-dimensional systems, the Berry holonomy corresponds to the Zak phase~\cite{Zak89}, which also reveals the winding number of the Hamiltonian mapping and has been measured in ultracold atoms in optical lattices~\cite{Atala2013}. The Berry connection, which is the foundation of the Berry phase, also serves as the base for building other topological indices for topological insulators and superconductors~\cite{Zhang_TIRev,Kane_TIRev}. 

At finite temperatures, a quantum statistical system is described by its density matrix, and a generalization of the Berry connection has been proposed by Uhlmann~\cite{Uhlmann,Uhlmann1,Uhlmann2}. Following the Uhlmann connection, the Uhlmann phase accumulated in a Uhlmann process reflects the Uhlmann holonomy of the purified state of the density matrix and serves as a topological index at finite temperatures~\cite{Viyuela14,Viyuela15,Viyuela2,Huang14}. By constructing the purified states of a two-level system on quantum computers, the Uhlmann phase has been measured~\cite{npj18}. Recently, the Uhlmann connection has been applied to time reversal invariant topological insulators \cite{Zhang21} and spin systems \cite{Galindo21,HouPRA21} to test topological behavior against finite-temperature effects. We also mention there are other pioneer works \cite{Sjoqvist00,Diehl,Diehl18} trying to generalize the concept of Berry connection or Berry phase of pure states to mixed states.

On the other hand, holonomic quantum computation~\cite{Zanardi99,Oreshkov09,Pachos_book,Carollo19} utilizes the phase factors of quantum wavefunctions, which may be viewed as a generalization of the holonomy, to achieve quantum computational operations. Protection of quantum information against local perturbations is expected when the phase from selected operations reflects a global change from the dynamics. There have been experimental demonstrations of holonomic quantum computation~\cite{Xu12,Feng13,Zhou17}. However, couplings to the environment may still affect some holonomic operations of the ground states~\cite{Guridi05}. To investigate  environmental effects on holonomy at finite temperatures, we will consider open-system descriptions of a system influenced by the environment. 

Several open quantum system approaches have been developed to handle system-environment couplings~\cite{Open-quantum-book,weiss2012quantum}. Time evolution of open quantum systems usually leads to mixed states. Because a full treatment of the dynamics of the system plus environment is usually complicated if not impossible, quantum master equations have been derived to emphasize the environmental effects on the system dynamics. Among the master equation approaches, the Lindblad equation~\cite{Lindblad76} has been a widely used method due to its simplicity while keeping the complete positivity during the dynamics. In the Lindblad equation, environmental effects manifest as quantum jumps. Some holonomic operations of the ground states of two-level systems are shown to be affected by certain types of quantum jumps~\cite{Guridi05}.  Since the Uhlmann holonomy is a generalization of the Berry holonomy to mixed quantum states, we will investigate its robustness against environmental effects. Ref.~\cite{Tidstrom03} shows the Uhlmann phase of a qubit can be influenced by decoherence modeled by the Lindblad equation, and here we will investigate several finite-temperature topological systems in Lindblad dynamics.

To be concrete, three model systems exhibiting topological properties in their ground states will be simulated and analyzed at finite temperatures, including the Su-Schrieffer-Heeger (SSH) model~\cite{SSH79}, Kitaev chain~\cite{Kitaev-chain}, and Bernevig-Hughes-Zhang (BHZ) model~\cite{BHZ-1}. The SSH model and Kitaev chain are 1D models while the BHZ model is a 2D time-reversal invariant topological insulator. In thermal equilibrium, the Uhlmann phases of the SSH model and Kitaev chain have been shown in Ref.~\cite{Viyuela14}, and the Uhlmann phase of the BHZ model has been discussed in Ref.~\cite{Zhang21}. All of them exhibit quantized Uhlmann phases and finite-temperature topological transitions.
We will show that the Uhlmann phase in the presence of certain types of Lindblad jump operators still remains quantized during the quantum dynamics, but other types of Lindblad jump operators may render the Uhlmann phase continuous. The demonstrations of protection of the Uhlmann holonomy against some environment-induced changes to the density matrix at finite temperatures offer hope for developing robust holonomy-based quantum operations beyond the ground states.

The rest of the paper is organized as follows.
Section~\ref{sec-Lind} summarizes a method for simulating and extracting the time-evolved density matrix of a quadratic quantum system following Lindblad dynamics. Section~\ref{sec-U} shows how to obtain the Uhlmann phase from purification of the density matrix. Sec.~\ref{sec-ex} shows the Uhlmann phase against different types of Lindblad jump operators from system-environment couplings of three examples: The SSH model, Kitaev chain, and BHZ model. The Uhlmann phase can remain quantized for selected Lindblad operators. Sec.~\ref{sec-conclu} concludes our study. Some details and additional results are given in the Appendix.

\section{Density matrix and Lindblad equation}\label{sec-Lind}
We briefly review the Lindblad formalism for describing time evolution of open quantum systems. The Hamiltonian of a general open quantum system can be expressed as
\be
H=H_S+H_B+H_I,
\ee
where $H_S$ is of the system we are interested in, $H_B$ describes the environment which may be thought of as a heat or particle bath. The last term describes the interactions between the system and environment. The initial state is postulated as a direct product of the density matrices of the system and environment as
$\rho(0)=\rho_S(0)\otimes\rho_B(0)$. The system plus the environment will follow the quantum unitary evolution given by
\be
\rho(t)=e^{i Ht}\rho_S(0)\otimes\rho_B(0) e^{-i Ht}.
\ee
We set $\hbar=1=k_B$. Since there are huge degrees of freedom in the environment, the full evolution is quite complicated. Meanwhile, one usually cares only about the properties of the system. Therefore, it is more convenient to trace out the degrees of freedom of the environment and obtain the evolution of the system as
\be
\rho_S(t)=\textrm{Tr}_B\Big[e^{i Ht}\rho_S(0)\otimes\rho_B(0) e^{-i Ht}\Big].
\ee
Although the initial state is a product state, the system and environment will become entangled due the interactions between them. An exact treatment of the above time evolution will still be quite intractable. At this stage, one usually introduces the so-called Born-Markov approximation and also rotation wave approximation to simplify the time evolution equation \cite{Open-quantum-book}. The final result can be written as a quantum master equation in the Lindblad form as
\be
i\frac{d\rho}{dt}=\mathcal{L}(\rho)=[\mathcal{H},\rho]+i\sum_{\mu}\Big(2L_{\mu}\rho L_{\mu}^{\dag}-\{L_{\mu}^{\dag}L_{\mu},\rho\}\Big).
\ee
Here $L_\mu$ are the Lindblad (quantum jump) operators, which encode the influence of the environmental dissipation effects on the system. Since the degrees of freedom of the environment have been traced out, $\rho$ here is the reduced density matrix of the system, and we have dropped the subscript of $\rho_S$.

We will mainly focus on noninteracting fermionic systems  described by the Hamiltonian
\be
\mathcal{H}=\sum_{ij}H_{ij}\dc_i c_j.
\ee
Here $H_{ij}$ is a Hermitian matrix describing the coefficients of the second quantized form. Moreover, we assume the Lindblad operators are linear in terms of the fermion operators:
\be
L^1_\mu=\sum_i l_{\mu i}c_i,~
L^2_\mu=\sum_i k_{\mu i}\dc_i.
\ee
Here we denote $L^1$ ($L^2$) as the loss (gain) dissipator.

For noninteracting fermions, the density matrix can be expressed in a Gaussian form as
\be
\rho=\frac 1Z e^{-\sum_{ij}\dc_i \Gamma_{ij} c_j},~
Z=\text{Tr}\left( e^{-\sum_{ij}\dc_i \Gamma_{ij}c_j}\right).
\ee
Thus, the density matrices cannot be expressed in a quadratic form of fermion operators. Because of this, it is not convenient to directly solve the Lindblad equation to find the density matrix. In order to obtain a solution, it is more desirable to convert the Lindblad equation to a matrix form.  To this end, we treat the correlation function $G_{ij}=\text{Tr}(\dc_i c_j\rho)$ as the central quantities to be evolved, instead of the density matrix. This method was suggested in
Ref.~\cite{Wang19} and also used in earlier works \cite{Bardyn_2013}. Making use of the Lindblad equation, the equation of motion for $G$ can be obtained as 
\be
&&\frac{d G(t)}{dt}=i[H^T,G(t)]-\{M_1^T+M_2, G(t)\}+2M_2\label{eq-G},\\
&&(M_1)_{ij}=\sum_\mu l^*_{\mu i}l_{\mu j},\quad
(M_2)_{ij}=\sum_\mu k^*_{\mu i}k_{\mu j}.
\ee
Here we introduce two Hermitian matrices $M_1$ and $M_2$, which reflect the loss and gain due to the environment, respectively. The non-zero gain dissipator gives rise to non-zero inhomogeneous terms in the equation of motion. Steady states are solutions of
$i[H^T,G(t)]-\{M_1^T+M_2, G(t)\}+2M_2=0$.
We will mainly concern the decay process around those possible steady states, meaning that we generally consider only the loss dissipation.

It is also convenient to rewrite the equation of motion as
\be
&&\frac{d G(t)}{dt}=i \Big[X G(t)-G(t)X^\dag\Big]-2M_2,\\
&&X=H^T+i(M_1^T+M_2).
\ee
Here $X$ may be considered as an effective Hamiltonian that takes account of the environmental dissipation. One can see that this effective Hamiltonian is generally non-Hermitian. With the assumption that all the dissipators are of the loss type, we can set $M_2=0$. If the Hamiltonian and quantum jump operators do not explicitly depend on time, the above equation can be integrated to give the time evolved correlation function
\be
G(t)=e^{i Xt}G(0)e^{-i X^\dag t}.
\ee
With a Gaussian type $\rho$, the correlation function can be computed and we obtain
\be
G(t)=\frac{1}{\exp[\Gamma(t)]+1}.
\ee
One can invert the above equation to find $\Gamma(t)$ at time $t$ from $G(t)$ as
\be
\Gamma(t)=\ln(G(t)^{-1}-1).
\ee
With $\Gamma(t)$ in hand, the density matrix $\rho$ at time $t$ is determined by
\be
\rho(t)=\frac{e^{\Gamma(t)}}{\textrm{Tr}\Big[e^{\Gamma(t)}\Big]}.
\ee
In the following, we will consider time evolution of selected one- or two- dimensional topological models under the influence of environmental dissipation. The initial state is taken to be the density matrix in thermal equilibrium:
\be
\rho(0)=\frac{e^{-\beta H}}{\textrm{Tr}\Big[e^{-\beta H}\Big]}.
\ee
Here $\beta=1/T$ is the inverse temperature. It is also interesting to consider a constant matrix as the initial-state density matrix, which then corresponds to the infinite-temperature state. Another possible choice is to consider a pure state at the starting point. In this case, however, $\rho$ is not a full-rank matrix. In all these cases, the results of time evolution are usually mixed states. Therefore, we need an indicator to detect the topological properties of the system. In the next section, we will describe the main topological tool based on the Uhlmann connection, which we will use throughout the paper.

\section{Uhlmann connection}
\label{sec-U}
We briefly review the Uhlmann connection, which is an extension of the Berry connection to finite temperatures~\cite{Uhlmann,Uhlmann1,Uhlmann2}. Firstly we recall the Berry connection is defined for a given eigenstate $\ket{\psi(r)}=e^{i\theta(r)}\ket{u(r)}$ with some parameter $r$ and some arbitrary phase factor $\theta(r)$. It helps to think of the Berry connection as a result of the parallel condition for two different states in the parameter space. This condition requires $\ep{\psi(r_1)|\psi(r_2)}>0$ in order for them to be parallel to each other~\cite{GPhase_book}. The infinitesimal version of the parallel-transport condition can be written as $\ep{\psi(r)|\p_r|\psi(r)}=0$, which gives rise to the Berry connection
\be
\p_r\theta=A_r=-i\ep{u|\p_r|u}.
\ee

At finite temperatures, we have to work with the density matrix $\rho$ of a mixed state instead of wave functions. By the spectral decomposition, we have $\rho=\sum_i p_i\ket{u_i}\bra{u_i}$ with eigenstates $\ket{u_i}$ and $i=1,\cdots,n$. In thermal equilibrium, $p_i$ is proportional to the Boltzmann weight and all eigenstates contribute to the density matrix. We introduce the amplitude decomposition, or purification, of the density matrix as
\be
\rho=ww^{\dagger},\qquad w=\sqrt{\rho}\,U.
\ee
Here $w$ may be thought of as the counterpart of the wave function for the mixed state. For a given $\rho$, $w$ is not uniquely determined because $wU$ with an arbitrary unitary matrix $U$ also gives rise to the same $\rho$. Similar to the case of wave functions, we can define the so-called Hilbert-Schmidt inner product $(w_1,w_2)\equiv\textrm{Tr}(w_1^\dag w_2)$ for two amplitudes \cite{GPhase_book}.

In order to define a connection in the amplitude space, we need a parallel condition. Only requiring $(w_1,w_2)>0$ is not strong enough to uniquely determine the $U(n)$ phase factor. Therefore,  Uhlmann \cite{Uhlmann} proposed the following parallel condition:
\be
w_1^{\dagger}w_2=w_2^{\dagger}w_1=C>0.
\ee
Here $C>0$ means that $C$ is a Hermitian and positive definite matrix. With this condition, the relative phase factor is uniquely defined.

Given two different amplitudes $w_1=\sqrt{\rho_1}U_1$ and $w_2=\sqrt{\rho_2}U_2$, the parallel condition provides us an expression 
$C^2=w_1^{\dagger}w_2w_2^{\dagger}w_1=U_1^{\dagger}\sqrt{\rho_1}\rho_2\sqrt{\rho_1}U_1$.
After taking the square root, one finds that
\be
C=U_1^{\dagger}\sqrt{\sqrt{\rho_1}\rho_2\sqrt{\rho_1}}\,U_1,
\ee
Combining the above equation with the parallel condition, we find a relative phase factor as
\be
U_2U_1^{\dagger}=\sqrt{\rho_2^{-1}}\sqrt{\rho_1^{-1}}\sqrt{\sqrt{\rho_1}\rho_2\sqrt{\rho_1}}.
\ee
This result can be considered as a finite version of the Uhlmann connection. Since $\rho^{-1}$ is used, the above formula requires the density matrix $\rho$ to be a full-rank matrix.

Like the Berry connection, it is more convenient to work with an infinitesimal Uhlmann connection. We can apply the above formula to $\rho_1=\rho$ and $\rho_2=\rho+\Delta k_\mu\p_\mu\rho$,  which are close in the $k_\mu$ parameter space. The parameter difference is $\Delta k_\mu$, and $\p_\mu=\frac{\p}{\p k_\mu}$. The infinitesimal Uhlmann connection has the form
\be
 A^U_\mu=\p_\mu U U^{\dagger},
\ee
which is an anti-Hermitian matrix. After some straightforward algebraic calculations, one can find an explicit expression of the Uhlmann connection as
\be
A^U_\mu&=&\ket{u_i}\bra{u_i}\frac{[\p_\mu\sqrt{\rho},\,\sqrt{\rho}]}{p_i+p_j}\ket{u_j}\bra{u_j}\nonumber \\
&=&\frac{(\sqrt{p_i}-\sqrt{p_j})^2}{p_i+p_j}\ket{u_i}\bra{u_i}\p_\mu\ket{u_j}\bra{u_j}.
\label{AU1}
\ee

The Uhlmann connection is a well defined $U(n)$ non-Abelian gauge field over the parameter space. Some features of the Uhlmann connection are worth mentioning. Since the above definition requires $\rho$ to be non-singular, Eq. (\ref{AU1}) cannot be directly applied to pure states. However, it has been shown \cite{Viyuela14} that in certain cases, the Uhlmann phase calculated from $A_U$ will approach the Berry phase as $T\to0$. A shortcoming of the Uhlmann connection is that the $U(n)$ bundle it is based on is topologically trivial, and all characteristic class, such as the Chern class and Chern character computed from the Uhlmann curvature, are all zero \cite{Diehl}. Ref. \cite{Viyuela2} proposed the so-called Uhlmann number that approaches the Chern number as $T\to0$,. There have been other modified Chern-number formulas to extract non-zero results from the Uhlmann connections~\cite{HeChern18}. However, those generalizations are not genuine topological indicators due to the triviality of the Uhlmann bundle. In contrast, the Uhlmann phase reflects the Uhlmann holonomy and remains valid for many topological models~\cite{Viyuela14,HouPRA21,Zhang21}

\section{Examples of Uhlmann phase against Lindblad dynamics}
\label{sec-ex}
Here we apply the Uhlmann connection to study the reaction of the Uhlmann holonomy of selected topological systems against quantum jumps during time evolution.
For one-dimensional system, we construct the Uhlmann-Wilson loop across the Brillouin zone:
\be
V=\mathcal{P}\exp\Big(\oint_C A^U_{\mu}d k_{\mu}\Big).
\ee
Here $C$ represents a closed loop as $k$ varies from $0$ to $2\pi$. Then we can define the Uhlmann phase as the phase angle of the inner product of the amplitudes $w(k=0)$ and $w(k=2\pi)$. Here we require $w(2\pi)$ to be parallel-transported from $w(0)$ according to Uhlmann's parallel condition. If we make use of the Uhlmann-Wilson loop defined above, the Uhlmann phase can be computed from
\be
\Phi^U &=&\arg\textrm{Tr}\ep{w(0)|w(2\pi)} \nonumber \\
&=&\arg\textrm{Tr}\Big[\rho_0\,\mathcal{P}\exp(\int_C A^U_\mu d k_\mu)\Big].
\ee
This index has already been used to study several 1D topological system at finite temperatures~\cite{Viyuela14}. 

For two-dimensional systems, one may cut the 2D Brillouin zone into many closed loops with fixed $k_y$ while $k_x$ varies from 0 to $2\pi$. Then the Uhlmann phase can be defined for each loop, allowing us to find a curve for evaluating $\Phi^U(k_y)$. The behavior of this curve indicates the topology of the system.
We caution that the procedures may fail to produce consistent results for some two-dimensional systems if the order of integration is reversed~\cite{Diehl}. If that happens, other types of finite-temperature indicators may be more suitable~\cite{Diehl18}. We remark that analytic calculations become quite unmanageable, as summarized in Appendix~\ref{app:analytics}, so we will resort to numerical calculations in the following.

\subsection{SSH model}
Our first example is the SSH model~\cite{SSH79}, a one-dimensional lattice hoping model with alternating hopping coefficients. In real space, its Hamiltonian is
\be
\mathcal{H}=\sum_j (w_1\dc_{1,j} c_{2,j+1}+w_2\dc_{2,j}c_{1,j+1})+H.c.
\ee
Here $j$ labels the unit cells and $1,2$ labels the two sites of a unit cell.
With periodic boundary condition, one can transform to momentum space and obtain
\be
&&\mathcal{H}=\sum_k\sum_{i,j}\dc_{i,k}H_{ij}(k)c_{j,k},\\
& &H=(w_1+w_2\cos k)\sigma_1+w_2\sin k\, \sigma_2. \nonumber
\ee
Here $\sigma_i$ with $i=1,2,3$ denotes the Pauli matrices. The initial state is the density matrix in thermal equilibrium with temperature $T$. It is convenient to express the Hamiltonian in terms of a 3D vector
\be
H=n_i(k)\sigma_i,\quad n_i=(w_1+w_2\cos k,\,0,\,w_2\sin k).
\ee
Here a summation is implicitly implied by repeated indices. Then the density matrix of the initial state at a given $k$ can be written as
\be
\rho(k)=\frac12\Big(1-\tanh(\frac{n}{T})\hat{n}_i(k)\sigma_i\Big).
\ee
Here $\hat{n}
_i=n_i/n$ and $n^2=\sum_j n_j^2$.

\begin{figure}
\centering
\includegraphics[width=0.8\columnwidth]{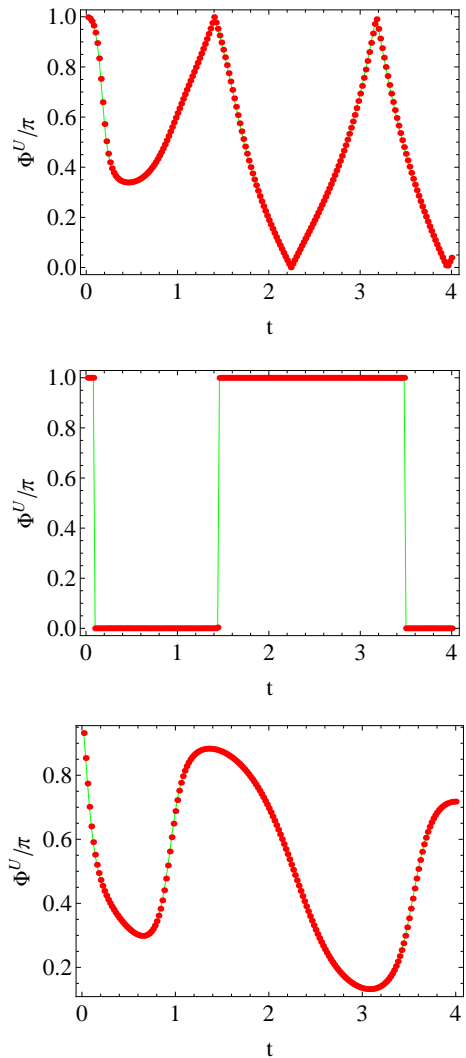}
\caption{The Uhlmann phase $\Phi^U$ of the SSH model as a function of time with $w_1=0.8$, $w_2=1$, $T=0.4$, and $\gamma^2=0.3$. The dissipation matrix has the form $M_1=\gamma^2(\sigma_0+\sigma_a)$. Top panel: $a=2$. Middle panel: $a=1$. Bottom panel: $a=3$.}
\label{SSH}
\end{figure}

The Lindblad dissipator $L_\mu$ will be chosen to be uniform, so we can compute the time-evolution in momentum space. Here, we will consider three types of Lindblad jump operators that correspond to three distinct types of dissipation matrices. The first one has the form
\be
L_j=\gamma(c_{j,1}-ic_{j,2}).
\label{L1}
\ee
Here $\gamma$ is real. Then we can construct the matrix $M$ in real space. After transforming to momentum space, we find
\be
M_1=\gamma^2(\sigma_0+\sigma_2),\quad M_2=0.
\ee
Here $\sigma_0$ is the 2 by 2 identity matrix. The effective Hamiltonian then becomes
\be
X=\left(
    \begin{array}{cc}
      \gamma^2 & w_1+\gamma^2+w_2 e^{-i k} \\
      w_1-\gamma^2+w_2 e^{i k} & \gamma^2
    \end{array}
  \right),
\ee
which is a non-Hermitian matrix with asymmetric hopping terms. 
We take a density matrix in thermal equilibrium at $T=0.4$ as the initial state and also choose the hopping coefficients to satisfy $w_2>w_1$. The Uhlmann phase of the initial state is $\Phi^U=\pm\pi$ (mod $2\pi$), which indeed indicates it is topological at this temperature. For the parameters used in Figure \ref{SSH}, the critical temperature of the SSH model is $T_c=0.48$, above which $\Phi^U=0$. In the top panel of Figure \ref{SSH}, we plot the numerical results of $\Phi^U$ as a function of time $t$. One can see that $\Phi^U$ varies with time in an almost periodic fashion. The Uhlmann phase deviates from $\pi$ for most of the time, indicating the system loses its quantized topological feature. At certain points of time, $\Phi^U$ comes back to the value of $\pi$, but it varies continuous in between.

Next, we consider a second choice of jump operators of the form
\be
L_j=\gamma(c_{j,1}+c_{j,2}).
\label{L2}
\ee
Here $\gamma$ is real.
Again, we find the $M$ matrix in momentum space as
\be
M_1=\gamma^2(\sigma_0+\sigma_1),\quad M_2=0.
\ee
The corresponding effective Hamiltonian
\be
X=\left(
    \begin{array}{cc}
      \gamma^2 & w_1+i\gamma^2+w_2 e^{-i k} \\
      w_1+i\gamma^2+w_2 e^{i k} & \gamma^2
    \end{array}
  \right)
\ee
is a non-Hermitian matrix with imaginary hopping coefficients. We take the same initial state as the previous case. The time evolution of $\Phi^U$ with this new set of jump operators is shown in the middle panel of Figure \ref{SSH}. When compared to the previous case, $\Phi^U$ is now always quantized at two values $0$ or $\pi$ (mod $2\pi$) during the dynamics. One can see that $\Phi^U$ jumps back and forth between these two quantized values as time evolves.

For the third trial, we choose $L_j=\gamma c_{j,1}$, which leads to the dissipation matrix $M_1=\gamma^2(\sigma_0+\sigma_3)$. The time evolution of the Uhlmann phase according to this type of Lindblad jump operators is shown in the bottom panel of Figure \ref{SSH}. One can see that $\Phi^U$ also oscillates without reaching the quantized values.

In summary, the above three examples of Lindblad jump operators give rise to the following dissipation matrix and effective Hamiltonian
\be
&&M_1=\gamma^2(\sigma_0+\sigma_a),\\
&&X=(w_1+w_2\cos k)\sigma_1+w_2\sin k\sigma_2+i \gamma^2 \sigma_a^T,
\ee
for $a=1,2,3$, and we have dropped the constant matrix in $X$. We see that $X$ is determined by the three coefficients in front of the Pauli matrices.
We observe that for $a=1$, only one of the three coefficients of $X$ contains a constant term, and $\Phi^U$ is quantized as time evolves. For the other two cases $a=2,3$, there are two coefficients containing constant terms, and the corresponding $\Phi^U$ is not quantized but varies continuously. Therefore, we conclude that $\Phi^U$ remains quantized during the dynamics only if the coefficients of the effective Hamiltonian $X$ contain at most one constant term in the coefficients.

As $T\rightarrow 0$, the Uhlmann phase in thermal equilibrium approaches the Berry phase for the SSH model~\cite{Viyuela14}. We found the results shown in Fig.~\ref{SSH} remain qualitatively the same as $T$ of the initial state decreases.
To check the generality of the results, we also consider possible initial states other than the thermal equilibrium state at finite $T$ and summarize the results in Appendix~\ref{app:init}.  The results suggest that the qualitative long-time behavior of the Uhlmann phase is the same regardless of the initial conditions.

\begin{figure}
\centering
\includegraphics[width=0.8\columnwidth]{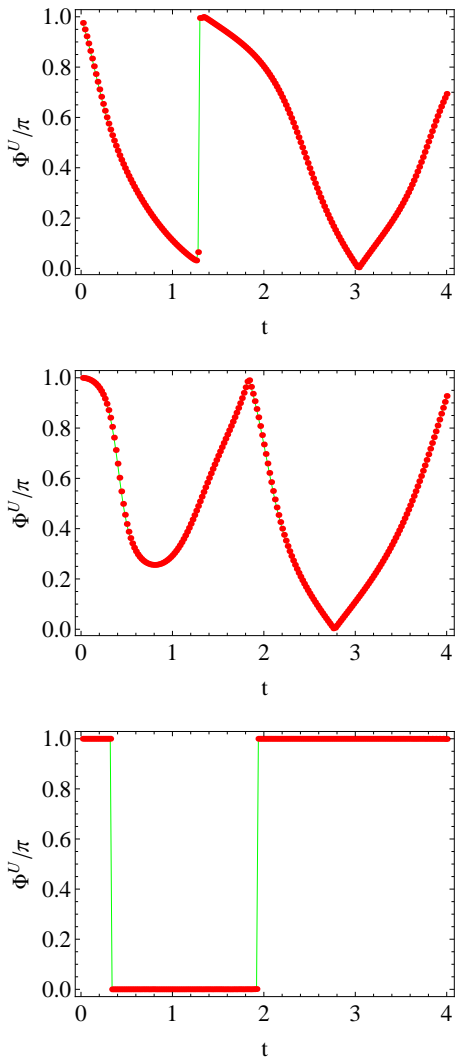}
\caption{The Uhlmann phase $\Phi^U$ of the Kitaev chain as a function of $t$. We assume that $w=1$, $\mu=0.5$, $\Delta=0.8$, $T=0.4$, and $\gamma^2=0.3$. The dissipation matrix has the form $M=\gamma^2(\sigma_0+\sigma_a)$. Top panel: $a=1$. Middle panel: $a=2$. Bottom panel: $a=3$. }
\label{K-chain}
\end{figure}

\subsection{The Kitaev chain}
As a second example of 1D topological systems, we turn to the Kitaev chain, which describes a one dimensional p-wave superconductor~\cite{Kitaev-chain}. In real space, its Hamiltonian is
\be
&&\mathcal{H}=\sum_i\Big[w(\dc_i c_{i+1}+\dc_{i+1}c_i)+\mu\dc_ic_i+\Delta(\dc_i\dc_{i+1}+c_{i+1}c_i)\Big]. \nonumber \\
&&
\ee
With periodic boundary condition, one can transform to momentum space and obtain the Hamiltonian
\be
& &\mathcal{H}=\sum_k\sum_{i,j}\psi^\dag_{i,k}H_{ij}(k)\psi_{j,k}, \\
& &H=(w\cos k+\mu)\sigma_3+\Delta\sin k\, \sigma_2. \nonumber
\ee
Here we define the two-component Nambu spinor $\psi_k=(c_{k},\,\dc_{-k})^T$. The Kitaev chain has a $\mathcal{Z}_2$ topological index \cite{Kitaev-chain}. At $T=0$, one can use the Berry phase $\Phi^B$ as the $\mathcal{Z}_2$ index. The topological regime exhibits $\Phi^B=\pm\pi$ (mod $2\pi$), which will occur when $|w|>|\mu|$.

Since the Kitaev chain contains pairing terms that mix particle- and hole- states, it cannot be written as $\sum_{ij}\dc_i H_{ij}c_j$. Therefore, we cannot use Eq. (\ref{eq-G}) to compute the time evolution. However, we notice that the two fermion-operators in the Numbu spinor $\psi_k$ may be considered to have different flavors of fermions. Since the fermion anti-commutation relations is symmetric about the creation and annihilation operators, we can make a particle-hole transformation with $\dc_{-k}\to c_{-k}$ for $k>0$ only. Then the Hamiltonian in momentum space is restored to the form as $\sum_{ij}\dc_i H_{ij}c_j$ after the transformation, which will be incorporated in the time evolution.

Similar to the discussion of the SSH model, we begin with the Lindblad jump operators of the form
\be
L_j=\gamma(c_{j}+\dc_{j}).
\ee
Here $j$ labels the lattice sites. Since the above jump operator is a combination of both creation and annihilation operators, Eq. (\ref{eq-G}) is not applicable in this case. Nevertheless, we can transform $L_j$ to momentum space and obtain
$L_k=\gamma(c_k+\dc_{-k})$. By applying the particle-hole transformation with $\dc_{-k}\to c_{-k}$ for $k>0$, we find that the Lindblad jump operator becomes $L_k=\gamma(c_k+c_{-k})$, which has the form of a pure lose dissipator. Thus, we can now use the method of Eq. (\ref{eq-G}) to find the dissipation matrix as
\be
M_1=\gamma^2(\sigma_0+\sigma_1),\quad M_2=0.
\ee
Again, we take a thermal-equilibrium state at $T=0.4$ as the initial state and choose the parameters to satisfy $w>\mu$, such that the initial state is topological. For the parameters used in Figure \ref{K-chain}, we find $T_c=0.58$. At $T=0.4$, we find $\Phi^U=\pi$ (mod $2\pi$) for the initial state, which indicates it is topological. The numerical results of $\Phi^U$ of the Kitaev chain as a function of time $t$ are plotted in the top panel of Figure \ref{K-chain}. As time evolves, one can see that $\Phi^U$ quickly decreases from $\pi$. But at certain time, it may abruptly jump back to the value $\pi$. Moreover, this type of behavior may repeat as time evolves.

For a comparison, we choose a different set of Lindblad jump operators of the form 
\be
L_j=\gamma(c_{j}+i\dc_{j}).
\ee
In momentum space, we find the dissipation matrices as
\be
M_1=\gamma^2(\sigma_0+\sigma_2),\quad M_2=0.
\ee
Starting with the same initial state, the time evolution of $\Phi^U$ is shown in the middle panel of Figure \ref{K-chain}. While $\Phi^U$ also oscillates in this case similar to the previous case, there is no abrupt jump in the current case. Moreover, $\phi^U$ reaches $\pi$ at some discrete points in time.

Finally, we test the form of Lindblad operators
\be
L_j=\gamma c_{j}
\ee
with real $\gamma$.
Then we find the dissipation matrices in momentum space satisfy
\be
M_1=\gamma^2(\sigma_0+\sigma_3),\quad M_2=0.
\ee
With the same initial state as the previous examples, we show the time evolution of $\Phi^U$ in the bottom panel of Figure \ref{K-chain}. In this case, $\Phi^U$ always takes the quantized values of $0$ and $\pi$ as time evolves. Therefore, there are also two types of behavior of the Kitaev chain, where $\Phi^U$ may vary continuously except some discrete jumps or it may remain quantized during the evolution, depending on the form of the Lindblad jump operators.

\begin{figure}
\centering
\includegraphics[width=0.8\columnwidth]{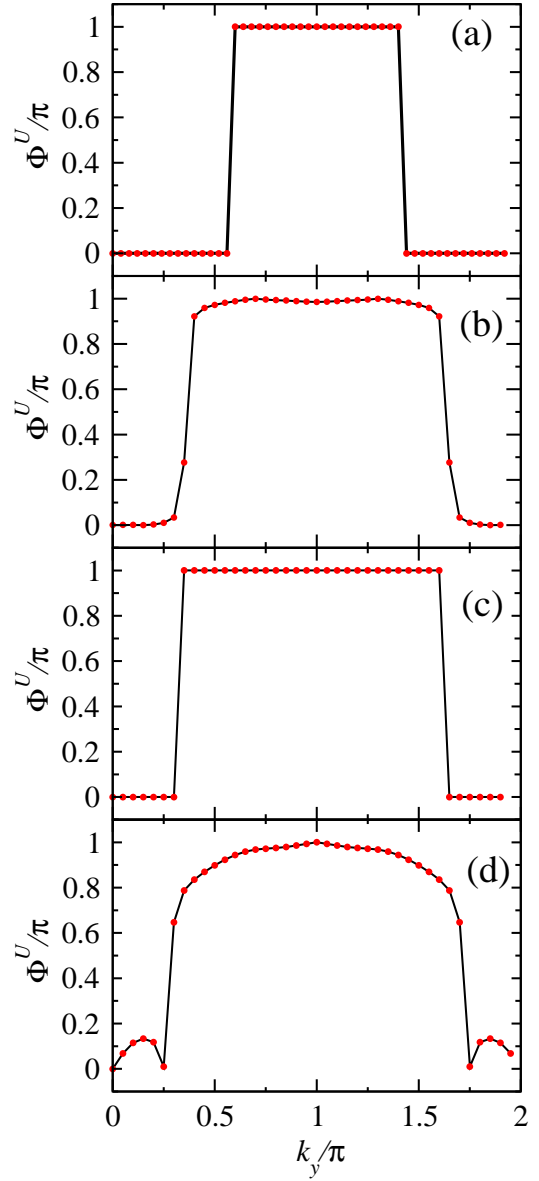}
\caption{The Uhlmann phase $\Phi^U$ of the BHZ model as a function of $k_y$. From top to bottom: The initial condition in thermal equilibrium with $T=0.4$ (a) and $\Phi^U$ at time $t=0.4$ of Lindblad dynamics with the dissipation matrices $M_1=\gamma^2(\sigma_0+\sigma_a)$ for (b) $a=1$, (c) $a=2$, and (d) $a=3$. We assume $m=0.8$, $\delta=0.2$, and $\gamma^2=0.3$.}
\label{BHZ}
\end{figure}

\subsection{The BHZ model}
Next, we consider a two-dimensional topological insulator with time reversal symmetry described by a prototypical example of the BHZ model. The Hamiltonian is given by
\be
H=\left(
    \begin{array}{cc}
      H_0(\vk) &  H_1 \\
      H_1^{\dag} & H_0^*(\vk)
    \end{array}
  \right).
\ee
The corresponding wave function is $\psi=(c_{1\uparrow},c_{2\uparrow},c_{1\downarrow},c_{2\downarrow})^T$, where the indices $i=1,2$ label the two orbitals and the arrows label the spins. Here $H_0$ is the Hamiltonian of the Qi-Wu-Zhang model \cite{Qi1} given by
\be
H_0=\sin k_x \sigma_1+\sin k_y \sigma_2+(m+\cos k_x+\cos k_y)\sigma_3,
\ee
where $\sigma_i$ for $i=1,2,3$ are the Pauli matrices acting on the space of the two orbitals.
The $H_1$ term is given by
\be
H_1=\left(
    \begin{array}{cc}
      0 &  \delta \\
      -\delta & 0
    \end{array}
  \right),
\ee
which describes the interacting between the up and down spins. The topology of the BHZ model can be capture by the Fu-Kane $Z_2$ index \cite{Fu-Z2}. Its topological regime appears when $\delta$ is small and $|m|<2$.

As discussed  previously \cite{Zhang21}, the topology of the BHZ model at finite $T$ can be indicated by the quantized Uhlmann phase $\Phi^U$. Importantly, the results of the BHZ model are robust against swapping the order of integration \cite{Zhang21} and justify the use of the Uhlmann phase for the BHZ model.
For the simulation of the BHZ model in Lindblad dynamics, the initial state is taken to be a thermal equilibrium state at temperature $T=0.4$, and we present $\Phi^U$ as a function of $k_y$ after integrating over $k_x$. $\Phi^U$ of the initial thermal-equilibrium state at $T=0.4$ is shown in Figure \ref{BHZ} (a). Here we assume $m=0.8$ and $\delta=0.2$, thus the BHZ model is topological at $T=0$. As shown in Fig.~\ref{BHZ} (a), the Uhlmann phase at $T=0.4$ with abrupt jumps of $\Phi^U$ from 0 to $\pi$ or vice versa as $k_y$ varies indicates the system remains topological. Ref.~\cite{Zhang21} shows the BHZ model becomes topologically trivial with only vanishing $\Phi^U$ when the temperature is above a critical value. For the parameters used in Fig.~\ref{BHZ}, we find $T_c=0.65$, above which $\Phi^U =0$ for any value of $k_y$ in thermal equilibrium.

We assume the same type of Lindblad jump operators as those for the SSH model and simulate the Lindblad dynamics of the BHZ model. In real space, the first type of Lindblad operators that we test can be written as
\be
L_j=\gamma(c_{j,1,s}+c_{j,2,s}), \quad \text{for}\quad s=\uparrow,\,\downarrow.
\ee
Transforming to momentum space, the corresponding dissipation matrices are given by
\be
M_1=\gamma^2\sigma_0\otimes(\sigma_0+\sigma_1),\quad M_2=0.
\label{M-BHZ}
\ee
Here the first $\sigma_0$ is acting on the spin space while the second one is acting on the orbital space. Now we present the numerical results of $\Phi^U$ of the final state of a short evolution time $t=0.4$ in Figure \ref{BHZ} (b). One can see that $\Phi^U$ is no longer quantized during time evolution, as the abrupt jumps of the equilibrium state become rounded after some time. These features indicate the topological properties fade away according to this type of Lindblad dynamics.

We may choose another type of Lindblad jump operators as
\be
L_j=\gamma(c_{j,1,s}-i c_{j,2,s}), \quad \text{for}\quad s=\uparrow,\,\downarrow.
\ee
Transforming to momentum space, we find that the dissipation matrices are given by
\be
M_1=\gamma^2\sigma_0\otimes(\sigma_0+\sigma_2),\quad M_2=0.
\label{M-BHZ-1}
\ee
With the same initial state, the $\Phi^U$ curve at $t=0.4$ is shown in Figure \ref{BHZ} (c). Importantly, $\Phi^U$ remains quantized after time evolution, showing that the topological state is robust against this type of Lindblad dissipators.

We also consider a third choice of the Lindblad jump operators with the form
\be
L_j=\gamma c_{j,1,s}, \quad \text{for}\quad s=\uparrow,\,\downarrow,
\ee
which correspond to the dissipation matrices in momentum space as
\be
M_1=\gamma^2\sigma_0\otimes(\sigma_0+\sigma_3),\quad M_2=0.
\ee
In this case, $\Phi^U$ is also not quantized, as shown in Fig.~\ref{BHZ} (d). The shape of $\Phi^U$ curve is similar to that of Eq. (\ref{M-BHZ}) except there are two small bumps near $k_y=0$ and $k_y=2\pi$. Similar to the SSH model and Kitaev chain, we only found one type of Lindblad jump operators that does not destroy the quantization of the Uhlmann phase of the BHZ model.

\subsection{Physical implications}
While the Berry connection leads to the Berry phase that can be measured in many physical systems~\cite{Zwanziger90,Bohm_book}, the Uhlmann connection and Uhlmann phase are elusive because of their root in purification of mixed states. Nevertheless, Ref.~\cite{npj18} shows the feasibility of simulating the purified states of a two-level system plus a reservoir in thermal equilibrium on a quantum computer and then extracting the Uhlmann phase from the purified state. The method is considered universal and has been proposed for simulating a general spin-$j$ system with a reservoir and extracting its Uhlmann phase on quantum computers~\cite{HouPRA21}. Moreover, Lindblad dynamics may be simulated efficiently on quantum computers~\cite{HuSR20,Kamakari22}. For the Uhlmann phase in Lindblad dynamics studied here, the simulation and extraction will be more challenging since the environment affect the dynamics, leading to interest time dependence of the Uhlmann phase of topological systems.

\section{conclusion}
\label{sec-conclu}
We have shown that depending on the types of Lindblad jump operators, the Uhlmann phase that characterizes finite-temperature topological properties of several systems may remain quantized or become continuous in Lindblad dynamics. The Lindblad equation and Uhlmann connection allow exact numerical solutions to characterize the reactions of the Uhlmann holonomy against various system-reservoir couplings modeled by the jump operators. The promising protocols for retaining the quantized Uhlmann phase at finite temperatures in the exemplary topological systems suggest viable extensions of quantum information technologies beyond the zero-temperature limit.

\begin{acknowledgments}
Y. H. was supported by the Natural Science Foundation of China under Grant No. 11874272 and Science Specialty Program of Sichuan University under Grant No. 2020SCUNL210. C. C. C. was supported by the National Science Foundation under Grant No. PHY-2011360.
\end{acknowledgments}

\appendix
\section{Challenges of analytical calculations}\label{app:analytics}
We attempt some analytical understanding of the time evolution of the SSH model described by the Lindblad equation. Firstly, the correlation function at the initial time is
\be
G(0)=\Big[\exp(\frac{n_i\sigma_i}{T})+1\Big]^{-1}.
\ee
Here we focus on the choice of $M_1=\gamma^2(\sigma_0+\sigma_2)$, and other cases can be discussed in a similar fashion,
The effective Hamiltonian $X$ is given by
\be
&&X=d_j\sigma_j+i\gamma^2\sigma_0,\\
&&d_1=n_1,\,d_2=-n_2-i\gamma^2,\,d_3=0. \nonumber
\ee
For this case, the time evolution operator becomes
\be
&&e^{-i X t}=e^{\gamma^2 t}\Big[\cos (d\cdot t)-i\sin (d\cdot t) (\hat{d}_j\sigma_j)\Big],\\
&&e^{i X^\dag t}=e^{\gamma^2 \cdot t}\Big[\cos (d^* t)+i\sin (d^*t) (\hat{d}_j^*\cdot \sigma_j)\Big].
\ee
Here $d^2=\sum_j d_j^2$ and $\hat{d}_j=d_j/d$.

It is more convenient to express the time evolution in terms of the inverse of the correlation functions as
\be
[G(t)]^{-1}=e^{i X^\dag t}[G(t)]^{-1} e^{-i X t}.
\label{inv-G}
\ee
Suppose that we parameterize the time-evolved $G(t)$ similar as $G(0)$, then
\be
G(t)=\Big[\exp(\frac{R_i\sigma_i}{T})+1\Big]^{-1}.
\ee
Substitute this form of correlation function into Eq. (\ref{inv-G}), we find the following equation for determining the vector $R_i$:
\be
&&\Big[\exp(\frac{R_i\sigma_i}{T})+1\Big]=e^{2\gamma^2 t}
\Big[\cos (d^* \cdot t)+i\sin (d^*\cdot t) (\hat{d}_j^*\sigma_j)\Big]\nonumber\\
&&\times\Big[\exp(\frac{n_i\sigma_i}{T})+1\Big]\cdot\Big[\cos (d\cdot t)-i\sin (d\cdot t) (\hat{d}_j\sigma_j)\Big].
\ee
Since both sides can be expanded by the Pauli matrices, we can in principle solve $R_i$ in terms of $n_i$, $d_i$ at time $t$. However, if a generic thermal state is used as the initial state, the above equation is too complicated to give a compact expression for the unknown vector $R_i$. Nevertheless, if we assume that $R_i$ is known, the Uhlmann connection can be computed from $R_i$ and has the expression
\be
A_U=-\frac{i}2f(R)\epsilon_{ijk}\hat{R}_id\hat{R}_j\sigma_k.
\ee
Here $\hat{R_i}=R_i/R$, $R^2=\sum_i R_i^2$ and $f(R)=1-\frac{1}{\cosh(R/T)}$.
Then the Uhlmann-Wilson loop can be determined by $A_U$ as
\be
V=\mathcal{P}\exp\Big(\oint_C A^U_{\mu}d k_{\mu}\Big).
\ee
Unfortunately, the operators $A^U$ along the integration path do not commute with each other, making the path-ordered integral intractable analytically.

\begin{figure}
\centering
\includegraphics[width=\columnwidth]{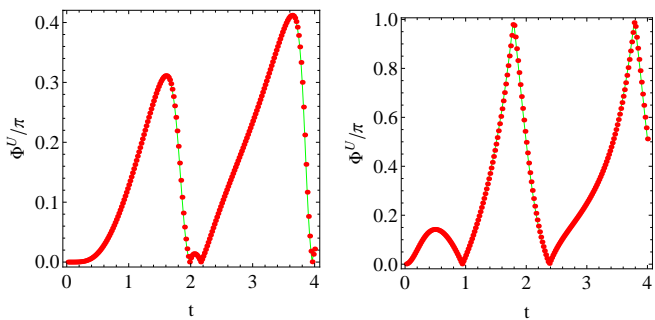}
\caption{The Uhlmann phase $\Phi^U$ of the SSH model as a function of $t$. The dissipation matrix is $M_1=\gamma^2(\sigma_0+\sigma_2)$. We take $w_1=0.8$, $w_2=1$, $T=0.4$, and $\gamma^2=0.9$. The left (or right) panel shows the case with the initial state of a complete disordered state (or an almost pure state). }
\label{SSH-1}
\end{figure}

\section{Robustness against initial conditions}\label{app:init}
To test the dependence on the initial condition, we try the completely disordered state, which can be thought of as the infinite-temperature state. We will use the SSH model to check the results. In this case, the density matrices are all constant matrices, and the Uhlmann connections are trivially vanishing. Thus the initial Uhlmann phase is zero. After time evolution, the density matrix starts to deviate from the constant matrix due to the environmental dissipation. The numerical results of $\phi^U$ of this case as a function of time are plotted in the left panel of Figure \ref{SSH-1}. One can see that $\Phi^U$ increases from zero to some finite values and also oscillates with time in a periodic fashion. Under our parameter choice, $\Phi^U$ cannot reach the non-trivial quantized value $\pi$.

While it is interesting to try an initial state that is a pure state, the initial density matrix then becomes singular, but the Uhlmann connection can only be defined for full-rank density matrices. To avoid this difficulty, we use a trial density matrix which is close to a pure state, given by $\rho(0)=\textrm{diag}(0.9,\,0.1)$. Since the trial density matrix has no $k$ dependence, the Uhlmann phase is trivial. The evolution of the Uhlmann phase with this initial condition is plotted in the right panel of Figure \ref{SSH-1}. 
As time increases, $\Phi^U$ keeps increasing and reaches $\pi$ at certain times. This is different from the case with a complete disordered state as the initial state. However, both results show continuous behavior of the Uhlmann phase as time evolves. 

\bibliographystyle{apsrev}

\end{document}